# Design and Analysis of Optimized Portfolios for Selected Sectors of the Indian Stock Market

Jaydip Sen
*Dept of Data Science*
*Praxis Business School*
Kolkata, INDIA
email: jaydip.sen@acm.org

Abhishek Dutta
*Dept of Data Science*
*Praxis Business School*
Kolkata, INDIA
email: duttaabhishek0601@gmail.com

*Abstract*— Portfolio optimization is a challenging problem that has attracted considerable attention and effort from researchers. The optimization of stock portfolios is a particularly hard problem since the stock prices are volatile and estimation of their future volatilities and values, in most cases, is very difficult, if not impossible. This work uses three ratios, the Sharpe ratio, the Sortino ratio, and the Calmar ratio, for designing the mean-variance optimized portfolios for six important sectors listed in the National Stock Exchange (NSE) of India. Three portfolios are designed for each sector maximizing the ratios based on the historical prices of the ten most important stocks of each sector from Jan 1. 2017 to Dec 31. 2020. The evaluation of the portfolios is done based on their cumulative returns over the test period from Jan 1, 2021, to Dec 31, 2021. The ratio that yields the maximum cumulative returns for both the training and the test periods for the majority of the sectors is identified. Additionally, the sectors which exhibit the maximum cumulative returns for the same ratio are also identified. The results provide useful insights for the investors in the stock market in making their investment decisions based on the current return and risks associated with the six sectors and their stocks.

## I. INTRODUCTION

Portfolio optimization is a challenging problem that has attracted considerable attention and effort from researchers. The optimization of stock portfolios is a particularly hard problem since the stock prices are volatile and estimation of their future volatilities and values, in most cases, is very difficult, if not impossible. The classical mean-variance optimization approach of portfolio optimization proposed by Markowitz is a theoretically elegant method with several practical limitations [1]. The most notable limitation is the unstable and inconsistent estimation of future returns of stocks computed based on their past returns. To address this shortcoming, maximization of some ratios (e.g., the Sharpe ratio) as the proxies for the expected returns is the approach that is generally followed in the mean-variance portfolio design.

This paper presents a step-by-step approach towards designing mean-variance optimization-based portfolios of stocks chosen from six important sectors listed in the National Stock Exchange (NSE) of India. Based on the NSE's report on Dec 31, 2020, the top ten stocks of each sector are first identified [2]. Three portfolios are built for each sector maximizing the Sharpe ratio, the Sortino ratio, and the Calmar ratio based on the historical stock prices from Jan 1, 2017, to Dec 31, 2020. The portfolios are tested over the period from Jan 1, 2021, to Dec 31, 2021, based on their cumulative returns. For each sector, the portfolio that yields the highest cumulative returns for the training and the test periods is identified. The ratio for which the portfolios yield the highest cumulative returns for the majority of the sectors for the training and the test periods is identified. For the mean-variance portfolio design approach, the ratio yielding the maximum cumulative return for the majority of the sectors is the one that should be maximized for designing the portfolios for the sectors. Moreover, the sectors for which the same portfolio has yielded the highest cumulative returns for both the training and the test periods are also identified. For these sectors, the investors will be able to maximize their profit using the same approach to the portfolio design.

The work has three unique contributions. First, the work compares the performances of the mean-variance portfolios designed for six sectors of the NSE on three ratios, the Sharpe ratio, the Sortino ratio, and the Calmar ratio, and identifies which ratio yields the highest cumulative returns for the portfolios for the majority of the sectors. The results give an idea about which ratio to use for maximizing the return, particularly for the Indian stock market. Second, the values of the ratios indicate the current return and risk of investments in the six sectors studied in the work. Finally, the sectors that exhibit the highest returns for the same ratio for both the training and the test periods of their portfolios provide opportunities for profitable investments for the investors.

The paper is organized as follows. Section II presents some of the existing approaches to portfolio design proposed in the literature. The research methodology followed in this work is described in Section III. Section IV the results and a detailed analysis of the observations. The paper is concluded in Section while identifying some future works.

## II. RELATED WORK

Portfolio design and optimization is a challenging problem that has attracted a lot of attention from the research community. Numerous approaches have been proposed to solve this complex problem involving robust stock price prediction and the formation of the optimized combination of stocks to maximize the return on investment. For future stock price prediction, the use of machine learning and deep learning algorithms and models have been proposed in many works [3-9]. The performances of the learning-based systems and models have been improved with the use of text mining and natural language processing algorithms on the rich and unstructured data on social media [10-13]. For estimating the future volatility and risk of stock portfolios the use of several variants of GARCH has been proposed in some works [14]. Several alternative methods and propositions to the classical mean-variance portfolio optimization also exist in the literature. The multiobjective optimization techniques [15], principal component analysis [16], deep learning LSTM models [17-19], future risk estimation methods [20], and swarm intelligence-based approaches [21-22] are some of the very popular portfolio optimization methods. Various other approaches such as the use of genetic algorithms [23], fuzzy sets [24], prospect theory [25], quantum evolutionary algorithms [26], and time series decomposition [27] for robust portfolio design are also proposed in the literature.

The present work evaluates the performance of the mean-variance approach to portfolio design on stocks chosen from some critical sectors of the Indian stock market. For each sector, portfolios are designed on three metrics, the Sharpe ratio, the Sortino ratio, and the Calmar ratio. The three portfolios for each sector are evaluated on their cumulative returns over the training and the test periods. The analysis of the results identifies the ratio that yields the highest cumulative returns for the majority of the sectors. Moreover, the cumulative returns yielded by the portfolios reveal the current profitability of the sectors. To the best of our knowledge, there is no such study existing in the current literature. Hence, the results of this work are expected to be useful to financial analysts and investors, especially those who are interested in the Indian stock market.

### III. DATA AND METHODOLOGY

This section discusses in detail the methodology followed in the current work especially focusing on the steps involved in designing the three portfolios for each of the six sectors. Three portfolio design approaches involve the maximization of three well-known ratios, the Sharpe ratio, the Sortino ratio, and the Calmar ratio. The methodology involves in following seven steps.

*A. Choice of the Sectors for Analysis*

Six important sectors are first chosen from the list of sectors in NSE so that they exhibit diversity in the Indian stock market. The sectors chosen in the present study are the following: *auto*, *banking*, *metal*, *fast moving consumer goods* (FMCG), *healthcare*, and *information technology* (IT). The NSE publishes monthly reports identifying ten stocks in each sector that have the most significant impact on the respective sector. Based on the NSE's report of Dec 31, 2020, the ten most critical stocks from each of the six sectors are identified [2].

*B. Extraction of Historical Stock Prices from the Web*

The *DataReader* function defined in the *pandas_datareader* module of Python is used to extract the historical prices of the stocks from the Yahoo Finance website for the period Jan 1, 2017, to Dec 31, 2021. The stock price data from Jan 1, 2016, to Dec 31, 2020, are used for building the portfolios on Dec 31, 2020. The performances of the portfolios are tested on the stock price records from Jan 1, 2021, to Dec 31, 2021. The *adjusted close* values are used for this univariate analysis.

*C. Designing the Maximum Sharpe Ratio Portfolios*

At this step, the portfolios with the maximum Sharpe ratio are designed. This involves several tasks which are discussed briefly in the following. For each stock in a given sector, the daily returns are computed as the change in the adjusted close values over successive days using the *pct_change* function of Python. The square root of the variance of the daily return values for a stock is computed to obtain the daily volatility values. The mean daily return and the mean daily volatilities are multiplied by a factor of 252 and the square root of 252 to obtain the annual return and the annual volatility figures. The covariance matrix of the daily returns of the stocks for a sector is now computed. The annual return and the annual risk for a portfolio are then computed. If a portfolio involves $n$ stocks and if $w_i$ represents the weight assigned to the stock $i$ which has an annual return of $R_i$, then the annual return ($R$) of the portfolio is: $R = \sum_{i=1}^{n} w_i * R_i$. The variance ($V$) of a portfolio is given by: $V = \sum_{i=1}^{n} w_i s_i^2 + 2 * \sum_{i,j} w_i * w_j * cov(i,j)$, where $s_i$ represents the annual standard deviation of the stock $i$, and $cov(i, j)$ is the covariance between the returns of stocks $i$ and $j$. The Sharpe ratio of a portfolio is computed as the ratio of the excess return the portfolio yields over the return of a risk-free portfolio to the risk of the current portfolio. Here, the portfolio annual standard deviation can be taken as a measure for the portfolio risk. To identify the portfolio with the maximum Sharpe ratio, 10,000 candidate portfolios are created in a *for loop* of Python allocating random weights to the stocks in the portfolios with a constrain that the sum of the weights is 1. These 10,000 portfolios are plotted on a two-dimensional space with the *x*-axis depicting the volatility and the *y*-axis the return. This plot, known as the *efficient frontier*, depicts the contour that includes portfolio points that yield the maximum return for a given risk or involve the minimum risk for a given return. While the left-most point on the efficient frontier identifies the portfolio with the lowest risk, the point for which the ratio of the return to the risk is maximized is the portfolio with the highest Sharpe ratio. The portfolio with the maximum Sharpe ratio is identified by using the Python function *idxmax* applied on the Sharpe ratios of all the 10,000 candidate portfolios. The portfolio with the maximum Sharpe ratio represents the portfolio with the optimum risk-adjusted return where the annual standard deviation is taken as the risk. The weights assigned by the Sharpe ratio maximizing portfolio to the stocks of each sector are noted.

*D. Designing the Maximum Sortino Ratio Portfolios*

The design of these portfolios differs from the Sharpe ratio maximizing portfolios in the way the risk of a portfolio is computed. While in the computation of the Sharpe ratio of a portfolio the risk is computed as the standard deviation of all return values, the Sortino ratio uses the standard deviation of the negative portfolio returns only in the risk computation. In other words, the Sortino ratio computation is based on the downside deviation of the return instead of its total standard deviation. Hence, the Sortino ratio is computed as the ratio of the result obtained by subtracting the return of a risk-free investment from the return of the current portfolio to the downside deviation of its return. Since this ratio focuses only on the negative side of the deviation of the returns from the mean return of a portfolio, it is sometimes assumed to serve as a better indicator of the risk as the positive deviations are benefits to the investors and they don't represent any risk. The steps involved in the design of the Sortino ratio maximizing portfolios are pretty similar to those used in building the Sharpe ratio maximizing portfolio except in the for loop in which the risks and the Sortino ratios of the portfolios are computed. For computing the downside deviation, the records with the negative returns are first identified and then the computation of their standard deviation and the Sortino ratios are carried out. Accordingly, the *x*-axis of the efficient frontiers plot depicts the downside deviation while the *y*-axis represents the return. The portfolio with the maximum Sortino ratio is identified for each sector and the weights assigned to the stocks by this portfolio are noted.

## E. Designing the Maximum Calmar Ratio Portfolios

The derivation of the Calmar ratio involves a different approach to the computation of the risk. The risk is computed as the maximum drawdown which is the maximum loss that is incurred as the return falls from a peak to a trough before attaining a peak again. The Calmar ratio is computed as the average annual rate of return of a portfolio to its maximum drawdown over the last three years. One of the advantages of using the Calmar ratio for evaluating portfolios is that it highlights the periodic and the cyclical fluctuations in their returns. However, since this ratio focuses only on the maximum drawdown, it provides the investors with a very narrow view of the risks involved. The efficient frontier consisting of the 10000 candidate portfolios for identifying the max Calmar ratio portfolio plots the drawdown along the x-axis while the y-axis depicts the annualized return. A python for loop is used that captures the running maximum drawdown which is updated only when the new drawdown exceeds its current value. At the end of the execution of the loop, the maximum drawdown over the entire period is identified. The annual return is then normalized by the max drawdown to derive the value of the Calmar ratio. Finally, the portfolio with the highest value of the Calmar ratio is identified. The weights assigned by the max Calmar ratio portfolio are noted.

## F. Visual Presentation of the Portfolios

Once the portfolios are designed for all the six sectors using the three approaches, they are presented in the form of pie diagrams, wherein the weights allocated to the stocks are depicted in percentage figures. For easier readability and better understanding, the weights are also presented in a tabular format using the properties of the pandas data frames in Python.

## G. Computation of the Portfolio Cumulative Returns

The weights associated with the stocks for all the six sectors under the three portfolio design approaches are now available. Based on the weights and the daily return values for the stocks, the cumulative returns for a given portfolio for the training and the test periods are computed as the weighted sum of the daily returns. The cumulative return values for three portfolios for the six sectors are plotted for the training and the test periods and their numeric values are presented in the form of tables. The portfolio yielding the highest cumulative return for the test period for a given sector is the best option for the investor among the three approaches. Moreover, since the portfolios are built on the training data, it is interesting to check which sectors exhibit the same portfolio yielding the maximum returns for both the training and the test dataset.

## IV. EXPERIMENTAL RESULTS

This section presents the detailed results of the performances of the portfolios and their analysis. The six sectors of the Indian stock market we choose are (i) *auto*, (ii) *banking*, (iii) *metal*, (iv) FMCG, (v) *healthcare*, and (vi) IT. The ratio-maximizing portfolios for each sector are implemented using Python 3.9.7 and its associated libraries of numpy, pandas, and matplotlib. The models are trained and validated on the GPU environment of Google Colab.

## A. Auto Sector

The ten most influential stocks of the auto sector with their weights that are used for deriving the auto sector index as reported by the NSE on Dec 31, 2020, are the following - Maruti Suzuki India (MSZ): 19.53, Tata Motors (TAM): 17.11, Mahindra and Mahindra (MHM): 15.85, Bajaj Auto (BAJ): 8.37, Eicher Motors (ECM): 7.15, Hero MotoCorp (HMC): 6.33, Balkrishna Industries (BAL): 3.73, Bharat Forge (BFG): 3.54, Ashok Leyland (ASL): 3.48, and Tube Investments of India (TII): 3.42 [2]. The stock of TII was first found to be listed in NSE on November 2, 2017, a date later than our "start date" from which we extracted the stock prices. Hence, the stock of TII is not included in the auto sector portfolio. Three portfolios are constructed maximizing the Sharpe ratio, the Sortino ratio, and the Calmar ratio, respectively, based on the daily stock returns during the training period. Fig 1(a) shows the efficient frontier of the portfolio that maximizes the Sharpe ratio. The portfolio with the lowest volatility is marked with the blue color, while the one with the maximum Sharpe ratio is marked with the red color. In Fig. 1(b), the portfolio with the lowest volatility among those which have negative deviance from the mean return value is marked with the blue color while the portfolio with the highest Sortino ratio is marked with the red color. Fig 1(c) depicts the portfolio with the highest Calmar ratio marked in the red color while the portfolio with the minimum drawdown value is marked with the blue color. The maximum values for the Sharpe ratio, the Sortino ratio, and the Calmar ratio are found to be 0.8417, 0.1970, and 0.1183, respectively.

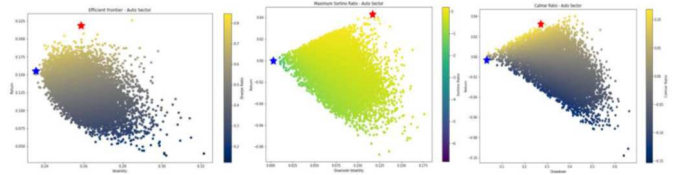

Fig. 1. Auto sector portfolio efficient frontiers: (a) the maximum Sharpe ratio, (b) the maximum Sortino ratio, and (c) the maximum Calmar ratio. The portfolios are built on the data from Jan 1, 2017, to Dec 31, 2020.

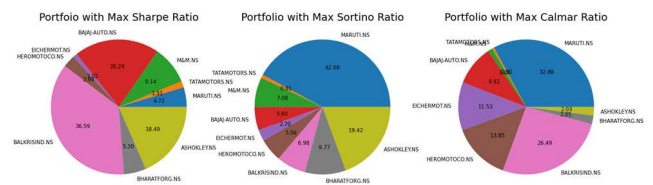

Fig. 2. Portfolio compositions for the auto sector: (a) The portfolio with the maximum Sharpe ratio, (b) the portfolio with the maximum Sortino ratio, and (c) the portfolio with the maximum Calmar ratio. The portfolios are built on the data from Jan 1, 2017, to Dec 31, 2020.

The compositions of three portfolios maximizing the ratios are shown in Fig 2. Table 1 presents the composition of the three portfolios for the auto sector in a tabular form. The stocks that received the maximum weights by the three portfolios are as follows: BAL (0.3659) by the max Sharpe Ratio portfolio, MSZ (0.4208) by the max Sortino ratio portfolio, and MSZ (0.3286) by the max Calmar ratio portfolio. The performances of the portfolios are evaluated by comparing the cumulative annual returns yielded by them for the training and the test period. Fig 3(a) and Fig 3(b) depict the cumulative returns of the portfolios for the training and the test period respectively. Table II presents the portfolio cumulative return values. The portfolio built on the maximum value of the Sharpe ratio is found to have yielded the highest cumulative return for both the training and the test periods.

The cumulative returns of the three portfolios for the auto sector for the in-sample (i.e. the training) data, and out-of-sample (i.e., the test) data are shown in Fig 3(a) and Fig 3(b), respectively.

TABLE I. PORTFOLIO COMPOSITION FOR AUTO SECTOR

| Stock | Max Sharpe | Max Sortino | Max Calmar |
|---|---|---|---|
| MSZ | 0.0472 | 0.4208 | 0.3286 |
| TAM | 0.0151 | 0.0081 | 0.0050 |
| MHM | 0.0914 | 0.0708 | 0.0108 |
| BAJ | 0.2029 | 0.0560 | 0.0961 |
| ECM | 0.0102 | 0.0270 | 0.1153 |
| HMC | 0.0294 | 0.0556 | 0.1385 |
| BAL | 0.3659 | 0.0698 | 0.2649 |
| BFG | 0.0530 | 0.0977 | 0.0205 |
| ASL | 0.1849 | 0.1942 | 0.0203 |

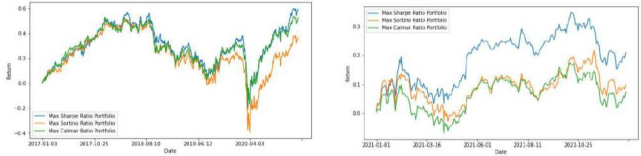

Fig. 3. The cumulative returns of the three portfolios of the auto sector: (a) the portfolio returns for the training samples and (b) the portfolio returns for the test samples.

TABLE II. CUMULATIVE PORTFOLIO RETURNS (AUTO SECTOR)

| Period | Max Sharpe | Max Sortino | Max Calmar |
|---|---|---|---|
| Training | 0.1518 | 0.0923 | 0.1341 |
| Test | 0.2126 | 0.1009 | 0.0743 |

*B. Banking Sector*

The top ten stocks of this sector and their weights are as follows: HDFC Bank (HDF): 27.80, ICICI Bank (ICI): 22.62, Kotak Mahindra Bank (KTB): 11.61, Axis Bank (AXS): 11.52, State Bank of India (STB): 11.45, IndusInd Bank (IIB): 5.91, AU Small Finance Bank (ASF): 2.33, Bandhan Bank (BNB): 1.75, Federal Bank (FDB): 1.70, and IDFC First Bank (IFB): 1.54 [2]. Two stocks, ASF and BNB, are not included in the portfolio analysis as the dates on which their prices were first available in the NSE were Jul 11, 2017, and Mar 27, 2018, respectively. The efficient frontiers of the three portfolios built for the banking sector are shown in Fig 4. These portfolios yielded the values of 0.9750, 1.0390, and 0.6000 for the maximum Sharpe ratio, Sortino ratio, and Calmar ratio, respectively.

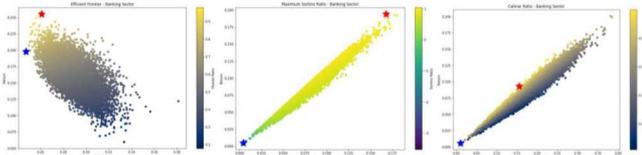

Fig. 4. Banking sector portfolio efficient frontiers: (a) the maximum Sharpe ratio, (b) the maximum Sortino ratio, and (c) the maximum Calmar ratio. The portfolios are built on stock prices from Jan 1, 2017, to Dec 31, 2020.

Fig 5 depicts the compositions of three portfolios for the banking sector. A tabular format of the portfolios is shown in Table III. It is found that all three portfolios allocated the largest weights to the stock KTB. The cumulative returns of the portfolios over the training and the test period are shown in Fig 6. It is observed in Table IV that the max Sharpe ratio portfolio has produced the maximum cumulative return during the training period, while for the test period, the portfolio with the max Calmar ratio has produced the highest return.

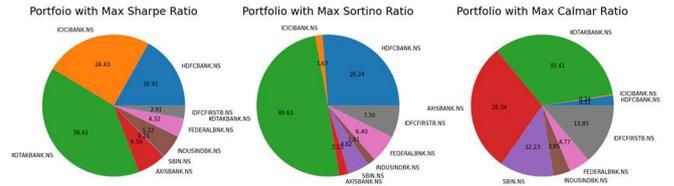

Fig. 5. Portfolio compositions for the banking sector: (a) the maximum Sharpe ratio, (b) the maximum Sortino ratio, and (c) the maximum Calmar ratio. The portfolios are built on the data from Jan 1, 2017, to Dec 31, 2020.

TABLE III. PORTFOLIO COMPOSITION FOR BANKING SECTOR

| Stock | Max Sharpe | Max Sortino | Max Calmar |
|---|---|---|---|
| HDF | 0.1691 | 0.2627 | 0.0231 |
| ICI | 0.2443 | 0.0167 | 0.0024 |
| KTB | 0.3941 | 0.4968 | 0.3341 |
| AXS | 0.0650 | 0.0213 | 0.2934 |
| STB | 0.0021 | 0.0483 | 0.1223 |
| IIB | 0.0532 | 0.0181 | 0.0385 |
| FDB | 0.0432 | 0.0641 | 0.0477 |
| IFB | 0.0291 | 0.0731 | 0.1385 |

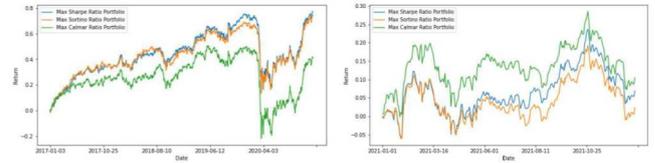

Fig. 6. The cumulative returns of the three portfolios of the banking sector: (a) the portfolio returns for the training samples and (b) the portfolio returns for the test samples.

TABLE IV. CUMULATIVE PORTFOLIO RETURNS (BANKING SECTOR)

| Period | Max Sharpe | Max Sortino | Max Calmar |
|---|---|---|---|
| Training | 0.1972 | 0.1890 | 0.1053 |
| Test | 0.0698 | 0.0239 | 0.1063 |

*C. Metal Sector*

The most critical stocks and their weights in this sector are Tata Steel (TAS): 20.45, Hindalco Ind (HND): 16.09, JSW Steel (JSS): 15.06, Adani Enterprises (ADE): 10.89, Vedanta (VED): 10.28, Coal India (COI): 7.09, Steel Authority of India (SAI): 3.59, Jindal Steel & Power [(JSP): 3.57, NMDC (NMD): 3.53, APL Apollo Tubes (AAT): 3.24 [2].

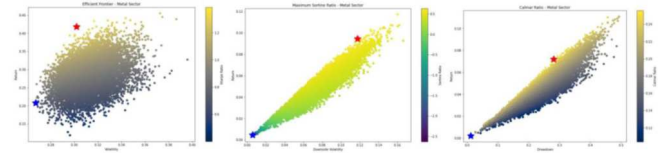

Fig. 7. Metal sector portfolio efficient frontiers: (a) the maximum Sharpe ratio, (b) the maximum Sortino ratio, and (c) the maximum Calmar ratio. The portfolios are built on the data from Jan 1, 2017, to Dec 31, 2020.

The efficient frontiers of the three portfolios which are designed based on the maximization of the Sharpe ratio, the Sortino ratio, and the Calmar ratio are exhibited in Fig 7. The maximum values of the three ratios yielded by the three portfolios are 1.3868, 0.6354, and 0.2558, respectively. The portfolio compositions of the three approaches are depicted in Fig 8 and also presented in a tabular form in Table V. It is found that while the max Sharpe ratio portfolio allocated the highest weight to the stock AAT, the other two portfolios assigned the stock JSS with the maximum weights. The cumulative of the three portfolios for the training and the test periods are shown in Fig 9. The results in Table VI show that the max Sharpe ratio portfolio has produced the highest cumulative returns for both training and test periods.

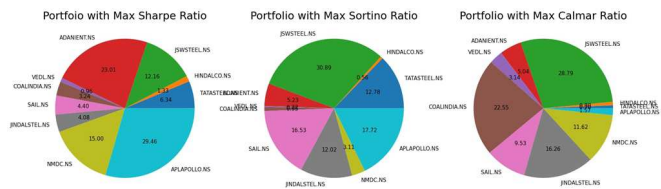

Fig. 8. Portfolio compositions for the metal sector: (a) the maximum Sharpe ratio, (b) the maximum Sortino ratio, and (c) the maximum Calmar ratio. The portfolios are built on the data from Jan 1, 2017, to Dec 31, 2020.

TABLE V. PORTFOLIO COMPOSITION FOR METAL SECTOR

| Stock | Max Sharpe | Max Sortino | Max Calmar |
|---|---|---|---|
| TAS | 0.0634 | 0.1278 | 0.0070 |
| HND | 0.0133 | 0.0056 | 0.0080 |
| JSS | 0.1216 | 0.3089 | 0.2879 |
| ADE | 0.2301 | 0.0523 | 0.0504 |
| VED | 0.0096 | 0.0022 | 0.0314 |
| COI | 0.0324 | 0.0095 | 0.2255 |
| SAI | 0.0440 | 0.1653 | 0.0953 |
| JSP | 0.0408 | 0.1202 | 0.1162 |
| NMD | 0.1500 | 0.0311 | 0.1162 |
| AAT | 0.2946 | 0.1772 | 0.0157 |

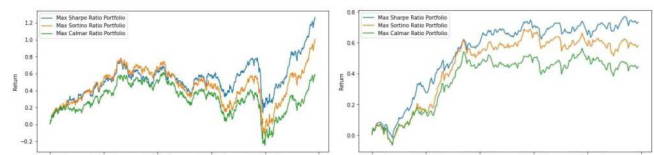

Fig. 9. The cumulative returns of the three portfolios of the metal sector: (a) the portfolio returns for the training samples and (b) the portfolio returns for the test samples.

TABLE VI. CUMULATIVE PORTFOLIO RETURNS (METAL SECTOR)

| Period | Max Sharpe | Max Sortino | Max Calmar |
|---|---|---|---|
| Training | 0.3230 | 0.2570 | 0.1504 |
| Test | 0.7462 | 0.5919 | 0.4519 |

*D. FMCG Sector*

The top ten stocks and their respective weights for the FMCG sector listed in the NSE are Hindustan Unilever (HIU): 27.62, ITC (ITC): 25.01, Nestle India (NSI): 9.22, Tata Consumer Products (TCP): 5.85, Britannia Industries (BRI): 5.59, Godrej Consumer Products (GCP): 4.81, Dabur India (DBI): 4.45, United Spirits (MCD): 3.52, Marico (MAR): 3.48, and Colgate Palmolive India (CPL): 2.59 [2]. The efficient frontiers of the three ratio maximizing portfolios are depicted in Fig 10. The maximum values of the Sharpe ratio, the Sortino ratio, and the Calmar ratio associated with the three portfolios are found to be 1.1906, 1.7995, and 1.3216, respectively. Fig 11 exhibits the compositions of the three portfolios while Table VII exhibits the weights allocated to the stocks by each portfolio in a tabular form. The stock NST received the highest allocation by the max Sharpe ratio portfolio while the other two portfolios assigned the maximum weights to the stock HUL. Fig 12 depicts the cumulative returns of the portfolios for the training and the test periods. The results presented in Table VIII elicits that the max Sortino ratio portfolio has produced the maximum cumulative return for the training period, while the max Calamar portfolio is found to be the most profitable one for the test period.

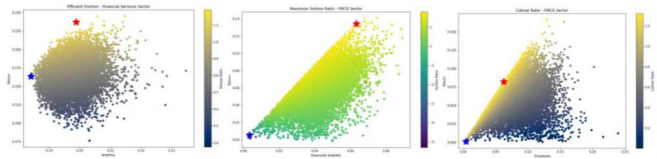

Fig. 10. FMCG sector portfolio efficient frontier: (a) the maximum Sharpe ratio, (b) the maximum Sortino ratio, and (c) the maximum Calmar ratio. The portfolios are built on data from Jan 1, 2017, to Dec 31, 2020.

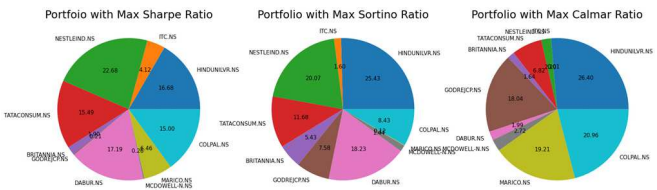

Fig. 11. Portfolio compositions for the FMCG sector: (a) the maximum Sharpe ratio, (b) the maximum Sortino ratio, and (c) the maximum Calmar ratio. The portfolios are built on the data from Jan 1, 2017, to Dec 31, 2020.

TABLE VII. PORTFOLIO COMPOSITION FOR FMCG SECTOR

| Stock | Max Sharpe | Max Sortino | Max Calmar |
|---|---|---|---|
| HIU | 0.1668 | 0.2543 | 0.2640 |
| ITC | 0.0412 | 0.0160 | 0.0001 |
| NSI | 0.2268 | 0.2007 | 0.0220 |
| TCP | 0.1549 | 0.1168 | 0.0682 |
| BRI | 0.0190 | 0.0543 | 0.0164 |
| GCP | 0.0021 | 0.0758 | 0.1804 |
| DBI | 0.1719 | 0.1823 | 0.0199 |
| MCD | 0.0026 | 0.0144 | 0.0272 |
| MAR | 0.0646 | 0.0012 | 0.1921 |
| CPL | 0.1500 | 0.0843 | 0.2096 |

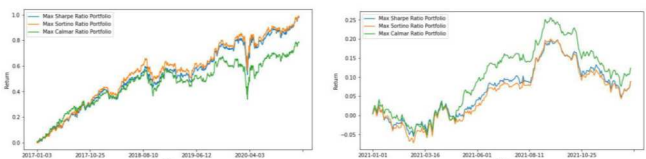

Fig. 12. The cumulative returns of the three portfolios of the FMCG sector: (a) the portfolio returns for the training samples and (b) the portfolio returns for the test samples.

TABLE VIII. CUMULATIVE PORTFOLIO RETURNS (FMCG SECTOR)

| Period | Max Sharpe | Max Sortino | Max Calmar |
|---|---|---|---|
| Training | 0.2497 | 0.2520 | 0.1997 |
| Test | 0.0903 | 0.0892 | 0.1257 |

*E. Healthcare Sector*

The top ten stocks and weights for the healthcare sector are Sun Pharmaceuticals (SUN): 18.22, Dr. Reddy's Laboratories (DRD): 11.90, Divi's Laboratories (DIV): 11.90, Apollo Hospitals Enterprise (APH): 10.07, Cipla (CIP): 9.73, Lupin (LPN): 4.57, Laurus Labs (LAU): 4.21, Aurobindo Pharma (AUP): 4.12, Biocon (BCN): 3.41, and Alkem Laboratories (AKL): 3.37 [2]. Fig 13 depicts the efficient frontiers of the three portfolios which are constructed based on maximizing the three ratios, Sharpe, Sortino, and Calmar. The maximum values for the three ratios yielded by the three portfolios are 1.4750, 1.3256, and 1.1352, respectively. The weight allocation done by the portfolios to the stocks is exhibited in Fig 14. Table IX shows the portfolio allocation in a tabular form. It is observed that while the max Sharpe ratio and the max Calmar ratio portfolios allocated the highest weights to the stock DVL, the max Sortino ratio portfolio assigned the stock BCN with the maximum weight. The cumulative returns produced by the three portfolios in the training and the test periods are exhibited in Fig 15. It is found from the results presented in Table X that the max Sharpe ratio portfolio has produced the maximum cumulative return for both the training and the test periods.

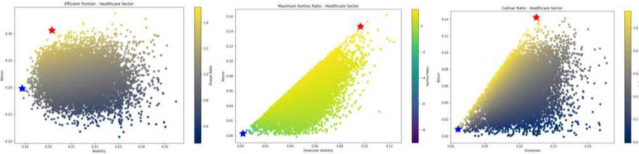

Fig. 13. Healthcare sector portfolios' efficient frontiers: (a) the maximum Sharpe ratio, (b) the maximum Sortino ratio, and (c) the maximum Calmar ratio. The portfolios are built on the data from Jan 1, 2017, to Dec 31, 2020.

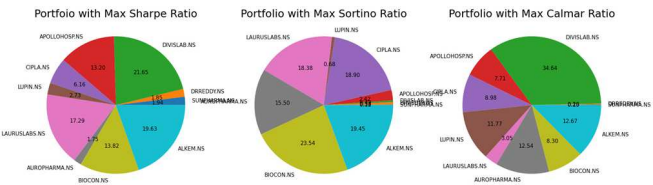

Fig. 14. Portfolio compositions for the healthcare sector: (a) the maximum Sharpe ratio, (b) the maximum Sortino ratio, and (c) the maximum Calmar ratio. The portfolios are built on the data from Jan 1, 2017, to Dec 31, 2020.

TABLE IX. PORTFOLIO COMPOSITION FOR HEALTHCARE SECTOR

| Stock | Max Sharpe | Max Sortino | Max Calmar |
|---|---|---|---|
| SUN | 0.0194 | 0.0012 | 0.0015 |
| DRD | 0.0185 | 0.0058 | 0.0020 |
| DIV | 0.2165 | 0.0042 | 0.3464 |
| APH | 0.1320 | 0.0242 | 0.0771 |
| CIP | 0.0616 | 0.1890 | 0.0898 |
| LPN | 0.0273 | 0.0068 | 0.1177 |
| LAU | 0.1729 | 0.1838 | 0.0305 |
| AUP | 0.0175 | 0.1550 | 0.1254 |
| BCN | 0.1382 | 0.2354 | 0.0830 |
| AKL | 0.1963 | 0.1945 | 0.1267 |

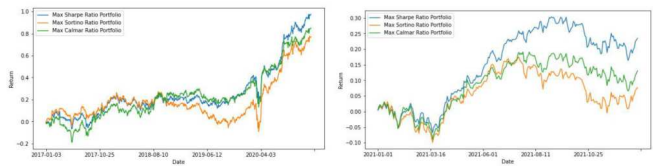

Fig. 15. The cumulative returns of the three portfolios of the healthcare sector: (a) the portfolio returns for the training samples and (b) the portfolio returns for the test samples.

TABLE X. CUMULATIVE PORTFOLIO RETURNS (HEALTHCARE SECTOR)

| Period | Max Sharpe | Max Sortino | Max Calmar |
|---|---|---|---|
| Training | 0.2489 | 0.1967 | 0.2174 |
| Test | 0.2388 | 0.0779 | 0.1327 |

*F. IT Sector*

The top ten stocks and weights for the IT sector are Infosys (INF): 27.52, Tata Consultancy Services (TCS): 25.09, HCL Technologies (HCL): 9.28, Wipro (WIP): 9.16, Tech Mahindra (TCM): 8.95, Larsen & Toubro Infotech (LTI): 5.34, MindTree (MNT): 4.91, MphasiS (MPS): 4.48, Coforge (CFG): 2.81, and L&T Technology Services (LTS): 2.46 [2]. The efficient frontiers of the three portfolios maximizing the Sharpe ratio, the Sortino ratio, and the Calmar ratio are exhibited in Fig 16. The maximum values of the three ratios yielded by the three portfolios are found to be 1.7835, 1.1970, and 0.8470, respectively. The compositions of the portfolios are exhibited in Fig 17 while the weight distribution made by the three portfolios to the ten stocks are presented in Table XI. The stocks LTI, HCL, and TCS received the maximum weight from the max Sharpe ratio, the max Sortino ratio, and the max Calmar ratio, respectively. The cumulative returns yielded by the three portfolios are shown in Fig 18. The results shown in Table XII indicate that the portfolio with the max Sharpe ratio has yielded the highest cumulative returns for both the training and the test periods.

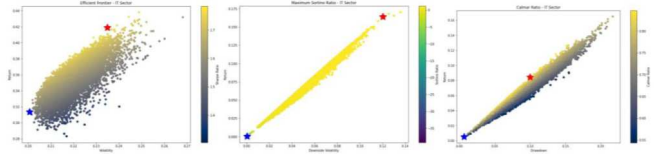

Fig. 16. IT sector portfolios' efficient frontiers: (a) the maximum Sharpe ratio, (b) the maximum Sortino ratio, and (c) the maximum Calmar ratio. The portfolios are built on data from Jan 1, 2017, to Dec 31, 2020.

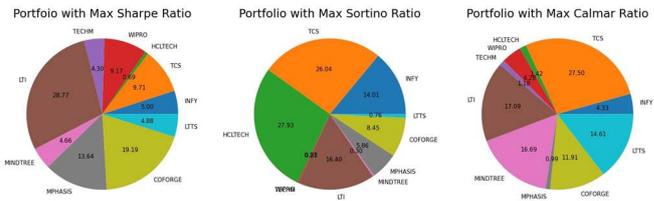

Fig. 17. Portfolio compositions for the IT sector: (a) the maximum Sharpe ratio, (b) the maximum Sortino ratio, and (c) the maximum Calmar ratio. The portfolios are built on the data from Jan 1, 2017, to Dec 31, 2020.

TABLE XI. PORTFOLIO COMPOSITION FOR IT SECTOR

| Stock | Max Sharpe | Max Sortino | Max Calmar |
|---|---|---|---|
| INF | 0.0500 | 0.1401 | 0.0433 |
| TCS | 0.0971 | 0.2604 | 0.2750 |
| HCL | 0.0069 | 0.2793 | 0.0142 |
| WIP | 0.0917 | 0.0021 | 0.0428 |
| TCM | 0.0430 | 0.0003 | 0.0118 |
| LTI | 0.2877 | 0.1640 | 0.1709 |
| MNT | 0.0466 | 0.0030 | 0.1669 |
| MPS | 0.1364 | 0.0586 | 0.0099 |
| CFG | 0.1919 | 0.0845 | 0.1191 |
| LTS | 0.0488 | 0.0076 | 0.1461 |

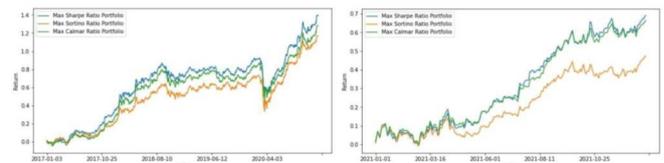

Fig. 18. The cumulative returns of the three portfolios of the IT sector: (a) the portfolio returns for the training samples and (b) the portfolio returns for the test samples.

TABLE XII. CUMULATIVE PORTFOLIO RETURNS (IT SECTOR)

| Period | Max Sharpe | Max Sortino | Max Calmar |
|---|---|---|---|
| Training | 0.3565 | 0.2988 | 0.3270 |
| Test | 0.7023 | 0.4812 | 0.6736 |

A summary of results is presented in Table XIII, in which for each sector, the portfolios with the maximum cumulative return for the training, and the test periods are mentioned. The maximum values of the three ratios for the portfolios are also mentioned. There are three observations in the results. First, it is observed that the portfolios built on maximizing the Sharpe ratio have yielded the highest returns for five among the six sectors. For the test data, the Sharpe ratio maximizing portfolios have yielded the maximum cumulative return for four sectors. The Sharpe ratio maximizing portfolios have outperformed their Sortino ratio and Calmar ratio-based counterparts. Second, there are four sectors, e.g., auto, metal, healthcare, and IT, for which the same portfolio (i.e., the Sharpe ratio maximizing portfolio) has performed the best for both training and test periods. Hence, for these four sectors, an investor who has designed the portfolios using the max Sharpe ratio approach will accrue the maximum profit over the test sample. For the other two sectors, however, the investors will have sub-optimal profits. The portfolios built on the training samples, i.e., the max Sharpe ratio portfolio for the banking sector, and the max Sortino ratio portfolio for the FMCG sector, will not yield the maximum returns. Third, the cumulative returns for the IT sector and the metal sector are the highest for the training and the test periods, respectively.

TABLE XIII. SUMMARY RESULTS FOR THE SIX SECTORS

| Sector | Highest Cum Return Port | | Max Sharpe Ratio | Max Sortino Ratio | Max Calmar Ratio |
|---|---|---|---|---|---|
| | *Trng* | *Test* | | | |
| Auto | Sharpe | Sharpe | 0.8417 | 0.1970 | 0.1183 |
| Banking | Sharpe | Calmar | 0.9795 | 1.0390 | 0.6000 |
| Metal | Sharpe | Sharpe | 1.3868 | 0.6354 | 0.2558 |
| FMCG | Sortino | Calmar | 1.1906 | 1.7995 | 1.3216 |
| HC | Sharpe | Sharpe | 1.4750 | 1.3256 | 1.1352 |
| IT | Sharpe | Sharpe | 1.7835 | 1.1970 | 0.8470 |

V. CONCLUSION

This work has presented three different approaches to portfolio design based on the maximization of three ratios – the Sharpe ratio, the Sortino ratio, and the Calmar ratio. Six critical sectors listed in the NSE of India are chosen, and portfolios are designed based on the three approaches for ten stocks from each sector on the historical stock prices from Jan 1, 2017, to Dec 31, 2020. The portfolios are evaluated on their cumulative returns for the period Jan1, 2021 to Dec 31, 2021. The portfolios built on the maximization of the Sharpe ratio are found to have yielded the maximum returns over the test period for four out of the six sectors. The study revealed the superiority of the Sharpe ratio as the metric. The future work includes a study of the other sectors listed in the NSE and other stock exchanges in the world using the three approaches of portfolio design.

ACKNOWLEDGMENT

The authors acknowledge the kind gesture of the Mae Fah Luang University, Thailand for allowing free registration in DASA'22 for the first author.

REFERENCES

[1] H. Markowitz, "Portfolio selection", *The Journal of Finance*, vol 7, no. 1, pp. 77-91, 1952, doi: 10.2307/2975974.

[2] NSE Website: http://www1.nseindia.com

[3] S. Mehtab and J. Sen, "Analysis and forecasting of financial time series using CNN and LSTM-based deep learning models," *Adv. in Dist. Comp. and Mach. Lrng.*, LNNS, vol 22, pp. 405-423, Springer, 2022, doi: 10.1007/978-981-16-4807-6_39.

[4] J. Sen and S. Mehtab, "Accurate stock price forecasting using robust and optimized deep learning models", *Proc. of CONIT'21*, Jun 2021, Hubballi, India, doi: 10.1109/CONIT51480.2021.9498565.

[5] S. Mehtab, J. Sen and A. Dutta, "Stock price prediction using machine learning and LSTM-based deep learning models," *Machine Learning and Metaheuristic Algorithms and Applications*, CCIS, vol 1366, pp 88-106, Springer, 2021, doi: 10.1007/978-981-16-0419-5_8.

[6] S. Mehtab and J. Sen, "A time series analysis-based stock price prediction using machine learning and deep learning models", *Int. J. of Bus. Forcstng. and Mktg. Int*, vol 6, no 4, pp. 272-335, 2021, doi: 10.1504/IJBFMI.2020.115691.

[7] J. Sen and T. Datta Chaudhuri, "A robust predictive model for stock price forecasting." *Proc. of BAICONF*, Dec 2017, Bangalore, India, doi: 10.36227/techrxiv.16778611.v1.

[8] C-C. Hung and Y-J. Chen, "Deep predictor for price movement from candlestick charts," *PloS ONE*, vol 16, no 6, doi: 10.1371/journal.pone.0252404.

[9] S. Mehtab and J. Sen, "Stock price prediction using CNN and LSTM-based deep learning models", *Proc. of DASA'20,* Nov 2020, Bahrain, doi: 10.1109/DASA51403.2020.9317207.

[10] Y. Li and Y. A. Pan, "A novel ensemble deep learning model for stock prediction based on stock prices and news." *Int. J. of Dat. Sci. and Anal.*, Sep 2021, doi: 10.1007/s41060-021-00279-9.

[11] S. Metab and J. Sen, "A robust predictive model for stock price prediction using deep learning and natural language processing", *Proc. of 7th BAICONF*, Dec 2019, doi: 10.36227/techrxiv.15023361.v1.

[12] M-L. Thormann, J. Farchmin, C. Weisser, R-M. Kruse, B. Safken, A. Silbersdorff, "Stock price predictions with LSTM neural networks and twitter sentiments", *Stat., Opt. & Info Compt.*, vol 9, no 2, pp. 268-287, Jun 2021, doi: 10.19139/soic-2310-5070-1202.

[13] Y. Zhang, J. Li, H. Wang and S-C. T. Choi, "Sentiment-guided adversarial learning for stock price prediction", *Front. Appl. Math. Stat.*, vol 7, 2021, doi: 10.3389/fams.2021.601105.

[14] J. Sen, S. Mehtab and A. Dutta, "Volatility modeling of stocks from selected sectors of the Indian economy using GARCH", *Proc of ASIANCON'21*, Aug 28-29, 2021, Pune, India, doi: 10.1109/ASIANCON51346.2021.9544977.

[15] Y. Zheng and J. Zheng, "A novel portfolio optimization model via combining multi-objective optimization and multi-attribute decision making", *Appld. Intl.*, 2021, doi: 10.1007/s10489-021-02747-y.

[16] J. Sen and S. Mehtab, "A comparative study of optimum risk portfolio and eigen portfolio on the Indian stock market," *Int. J. of Bus. Forcstng. and Mktg. Intl.*, vol 7, no 2, pp. 143-195, 2022, doi: 10.1504/IJBFMI.2021.120155.

[17] J. Sen, S. Mondal and G. Nath, "Robust portfolio design and stock price prediction using an optimized LSTM model", *Proc. of IEEE INDICON'21*, pp 1-6, Dec 2021, Guwahati, India, doi: 10.1109/INDICON52576.2021.9691583.

[18] J. Sen, A, Dutta and S. Mehtab, "Stock portfolio optimization using a deep learning LSTM model," *Proc. of Mysurucon*, October, Hassan, India, doi: 10.1109/MysuruCon52639.2021.9641662.

[19] J. Sen, S. Mehtab, A. Dutta and S. Mondal, "Precise stock price prediction for optimized portfolio design using an LSTM model", *Proc. of OCIT*, December, Bhubaneswar, India, doi: 10.1109/OCIT53463.2021.00050.

[20] J. Sen, S. Mehtab, A. Dutta and S. Mondal, "Hierarchical risk parity and minimum variance portfolio design on NIFTY 50 stocks", *Proc. of DASA*, Dec 2021, Bahrain, doi: 10.1109/DASA53625.2021.9681925.

[21] M. Corazza, G. di Tollo, G. Fasano and R. Pesenti, "A novel hybrid PSO-based metaheuristic for costly portfolio selection problem", *Ann Oper Res*, vol 304, pp. 109-137, 2021, doi: 10.1007/s10479-021-04075-3.

[22] A. Thakkar and K. Chaudhuri, "A comprehensive survey on portfolio optimization, stock price and trend prediction using particle swarm optimization", *Arch.Comput Methd in Engg.*, vol 28, pp 2133-2164, doi: 10.1007/s11831-020-09448-8.

[23] M. Kaucic, M. Moradi and M. Mirzazadeh, "Portfolio optimization by improved NSGA-II and SPEA 2 based on different risk measures", *Fin. Innv.*, vol 5, no 26, 2019, doi: 10.1186/s40854-019-0140-6.

[24] M. Karimi, H. Tahayori, K. Tirdad and A. Sadeghian, "A perceptual computer for hierarchical portfolio selection based on interval type-2 fuzzy sets", *Granul. Comp.*, 2022, doi: 10.1007/s41066-021-00311-0.

[25] Y. Li, B. Zhou and Y. Tan, "Portfolio optimization model with uncertain returns based on prospect theory", *Complex Intell. Syst.*, 2021, doi: 10.1007/s40747-021-00493-9.

[26] Y-H. Chou, Y-C. Jiang and S-Y. Kuo, "Portfolio optimization in both long and short selling trading using trend ratios and quantum-inspired evolutionary algorithms", *IEEE Access*, vol 9, pp. 152115-152130, 2021, doi: 10.1109/ACCESS.2021.3126652.

[27] J. Sen and T. Datta Chaudhuri, "Understanding the sectors of Indian economy for portfolio choice", *Int. J. of Bus. Forcstng. and Mktg. Intl.*, vol 4, no 2, pp. 178-222, 2018, doi: 10.1504/IJBFMI.2018.090914.